\documentclass[reprint,amsmath,amssymb,aps]{revtex4-2}

\usepackage{graphicx}
\usepackage{dcolumn}
\usepackage{bm}
\usepackage{url}
\usepackage{here}
\begin{document}
\title{
Pseudogap formation in organic superconductors
}

\author{Shusaku Imajo$^{1}$}
\email{imajo@issp.u-tokyo.ac.jp}
\author{Takuya Kobayashi$^{2}$}
\author{Yuki Matsumura$^{3}$}
\author{Taiki Maeda$^{3}$}
\author{Yasuhiro Nakazawa$^{3}$}
\author{Hiromi Taniguchi$^{2}$}
\author{Koichi Kindo$^{1}$}

\affiliation{
$^1$Institute for Solid State Physics, University of Tokyo, Kashiwa, Chiba 277-8581, Japan\\
$^2$Graduate School of Science and Engineering, Saitama University, Saitama 338-8570, Japan\\
$^3$Graduate School of Science, Osaka University, Osaka 560-0043, Japan
}

\date{\today}

\begin{abstract}
The condensation of paired fermions into superfluid states changes progressively depending on the coupling strength.
At the midpoint of the crossover between Bardeen--Cooper--Schrieffer (BCS) weak-coupling and Bose--Einstein condensate (BEC) strong-coupling limits, paired fermions condensate most robustly, thereby leading to the emergence of a pseudogap due to enhanced pairing fluctuations.
In the case of electrons in solids, excessively strong interactions often induce competing electronic orders instead of strong-coupling superconductivity, and experimental comprehension of the pseudogap remains incomplete.
In this study, we provide experimental evidence demonstrating the opening of a pseudogap, marking the incipient stage of the BCS-BEC crossover in the organic system $\kappa$-(BEDT-TTF)$_2$$X$.
By controlling electron correlations, we investigate the thermodynamic properties of the BCS-BEC crossover and pseudogap phase.
Since the superconductivity of $\kappa$-(BEDT-TTF)$_2$$X$ arises from a simple Fermi liquid that does not exhibit any other electronic orders, our study shed light on the inherent nature of the BCS-BEC crossover.
\end{abstract}

\maketitle
According to the well-established Bardeen--Cooper--Schrieffer (BCS) theory, fermions undergo the formation of weakly coupled Cooper pairs below a critical temperature $T_{\rm c}$.
On the other hand, Bose--Einstein condensate (BEC) is a quantum phenomenon observed in bosonic systems, where bosons condensate into the lowest-energy state as a result of the overlap of their wave functions at low temperatures.
When the pairing interactions between fermions are enhanced and the pairing deviates from the weak-coupling BCS regime, paired fermions can be regarded as a boson.
In such case, fermion pairing into bosons occurs at $T^{\ast}$ (greater than $T_{\rm c}$), and subsequently, depending on the phase stiffness of the superfluid, the system exhibits the BEC state below $T_{\rm c}$.
The continuous crossover phenomenon from the weak-coupling BCS regime to the strong-coupling BEC regime is known as the BCS-BEC crossover\cite{Eagles1969,Drechsler1992,Melo1993,Jochim2003,Zwierlein2003,Gaebler2010,Randeria2014,Chen2022}.
Experimental investigations of the BCS-BEC crossover have been predominantly conducted in ultracold atomic system, where pairing interactions in a Fermi gas can be controlled through Feshbach resonance\cite{Jochim2003,Zwierlein2003,Gaebler2010,Randeria2014}. 
For the BCS-BEC crossover of superconductivity in solids, its behavior is modified due to the difference of the kinetic energy associated with the electron motion in a underlying crystal lattice\cite{Chen2022}.
However, its experimental understanding remains limited due to the difficulty for the significant enhancement of interactions for electron pairing.
\begin{figure}
\begin{center}
\includegraphics[width=\hsize]{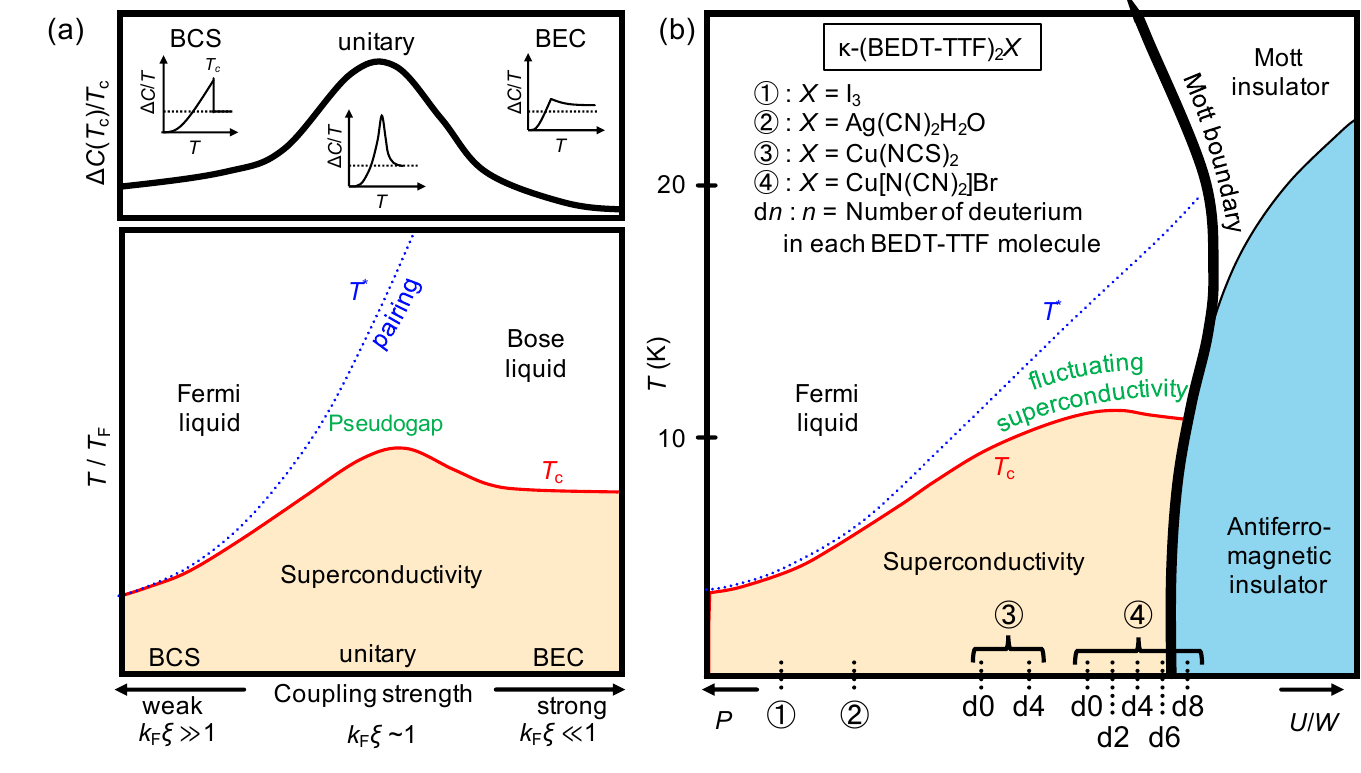}
\end{center}
\caption{
(a) Bottom panel: Schematic of pairing temperature $T^{\ast}$ (blue dotted curve) and condensation temperature $T_{\rm c}$ (red solid curve), both reduced by the Fermi temperature ($T_{\rm F}$) as a function of dimensionless coupling strength $k_{\rm F}$$\xi$.
At the unitary limit, where $k_{\rm F}$$\xi$ is of order unity, a pseudogap opens in the temperature range between $T_{\rm c}$ and $T^{\ast}$.
Upper panel: Coupling strength dependence of heat capacity jump at $T_{\rm c}$, with schematic $\Delta$$C$/$T$ versus $T$ for each limit in the insets.
(b) Electronic phase diagram of the half-filled $\kappa$-(BEDT-TTF)$_2$$X$.
The $\kappa$-type salts with different counter anions $X$ are positioned at different positions on the horizontal axis characterized by pressure/electron correlations.
The number $n$ in the expression "d$n$" stands for the number of deuterium atoms in ethlene group of each BEDT-TTF molecule.
}
\label{fig1}
\end{figure}

Since the BCS-BEC crossover is governed by whether fermion pairs can be viewed as bosons, the spatial size of fermion pairs and the average interfermion distance are principal parameters.
In the case of superconducting Cooper pairs, these quantities correspond to the coherence length $\xi$ and the average interelectron distance 1/$k_{\rm F}$ ($k_{\rm F}$ denotes the Fermi wave number), respectively.
The bottom panel of Fig.~\ref{fig1}a \cite{Melo1993,Randeria2014,Chen2022} depicts a schematic phase diagram illustrating the basic behavior of the BCS-BEC crossover.
It is important to note that the details of the crossover behavior vary depending on system parameters, such as pairing symmetry and dimensionality\cite{Chen1999}.
On the BCS side, characterized by a large pair size, the relation $\xi$ $\gg$1/$k_{\rm F}$ holds.
In contrast, on the BEC side, where tightly bound fermions are considered as local bosons, the relation $\xi$ $\ll$1/$k_{\rm F}$ applies.
\begin{figure*}
\begin{center}
\includegraphics[width=\hsize]{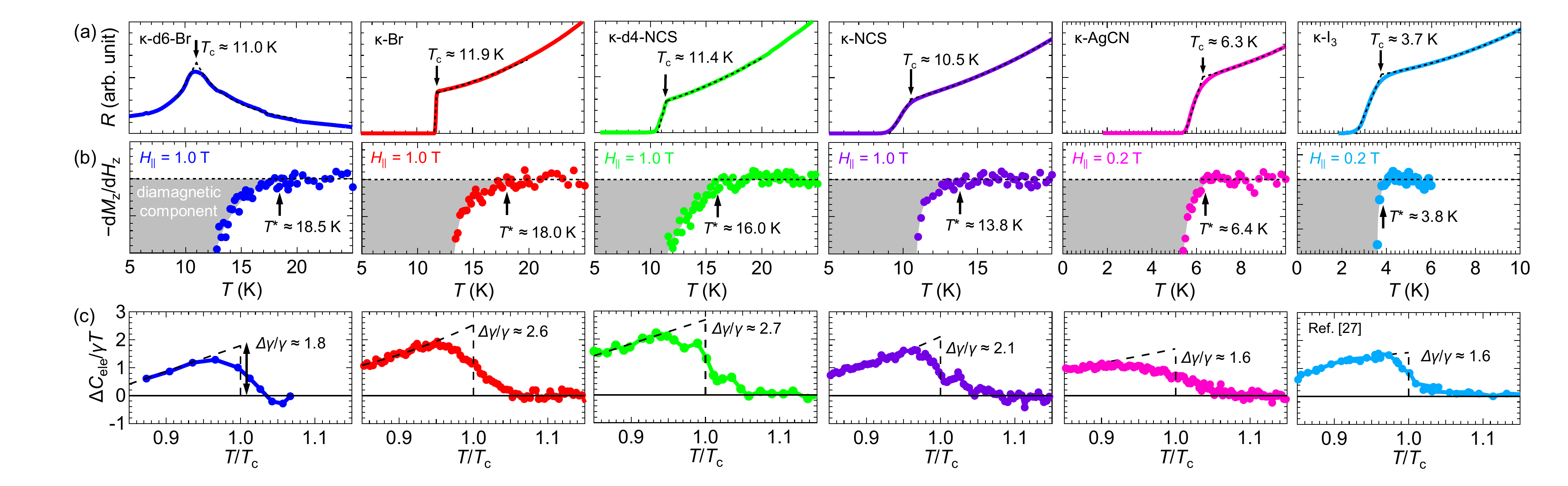}
\end{center}
\caption{
(a) Resistivity $R$ and (b) diamagnetic susceptibility d$M_z$/d$H_z$ as a function of temperature.
The black arrows in (a) and (b) indicate $T_{\rm c}$ and $T^{\ast}$, respectively.
The gray shaded areas in (b) represent the diamagnetic component.
(c) Heat capacity jump $\Delta$$C_{\rm ele}$/$\gamma$$T$ versus reduced temperature $T$/$T_{\rm c}$.
The data of $\kappa$-I$_{3}$ is taken from Ref. \cite{Wosnitza1994}.
The dashed curves are the estimation of mean-field behavior to evaluate the heat capacity jump at $T_{\rm c}$, $\Delta$$\gamma$/$\gamma$.
}
\label{fig2}
\end{figure*}
In the vicinity of the crossover region, $\xi$$\sim$1/$k_{\rm F}$ ($k_{\rm F}$$\xi$ $\sim$1), referred to as the unitary regime, the pairs robustly condensate.
This results in a peak structure in the heat capacity jump at $T_{\rm c}$ as a function of $k_{\rm F}$$\xi$, as shown in the upper panel of Fig.~\ref{fig1}a \cite{Zwierlein2003,Haussmann2007,Harrison2022}.
Even above $T_{\rm c}$, the presence of superfluid fluctuations lead to the preformation of Cooper pairs at a pairing temperature $T^{\ast}$, which suppress low-energy single-particle excitations.
This is the manifestation of a pseudogap.
Various superconductors have been investigated as potential candidates for exploring the BCS-BEC crossover regime \cite{Harrison2022,Timusk1999,Kasahara2016,Cao2018,Nakagawa2021,Suzuki2022}.
However, in low-carrier density systems possesing strong electron correlations, other electronic orderings or critical phenomena often occur above $T_{\rm c}$.
Consequently, superconductivity in these systems does not arise from a simple Fermi liquid.
To observe a pseudogap, it is essential to observe the suppression of single-particle excitations in the normal state.
Therefore, systems with complex electronic states are not ideal for studying the physics of the pseudogap.
This difficulty in accurately detecting the pseudogap has hindered detailed discussions regarding the properties of the BCS-BEC in solid-state systems.

Half-filled $\kappa$-type organic superconductors are strongly correlated systems that exhibit a superconducting transition from a simple Fermi liquid, as shown in Fig.~\ref{fig1}b \cite{Kanoda2006,Nakazawa2018}.
When the electron correlation $U$/$W$ is small (or pressure $P$ is high), the Fermi liquid shows a superconducting transition at low temperatures.
An increase in $U$/$W$ induces band renormalization, leading to enhancement of the effective mass of electrons and $T_{\rm c}$.
Further increasing $U$/$W$ causes the ground state to discontinuously become the Mott antiferromagnetic insulator across the first-order Mott phase transition \cite{Kanoda2006,Nakazawa2018}.
Near the Mott boundary, the value of $T_{\rm c}$ becomes the maximum, and fluctuating superconductivity was observed in a relatively wide temperature range even above $T_{\rm c}$ \cite{Nam2007,Tsuchiya2012,Tsuchiya2013}.
Since these prior studies investigated only strongly correlated (high $U$/$W$) salts, the fluctuating superconductivity has been discussed from a perspective of Coulomb penalty associated with the proximity to the Mott boundary.
A recent NMR study\cite{Furukawa2023} suggests that the predominant origin of the superconducting fluctuations should be attributable to pseudogap formation.
Nevertheless, experimental reports of less correlated (low $U$/$W$) salts have been lacking to date, and therefore, it has been challenging to ascertain the origin of the fluctuating superconductivity.

In this study, we report that in the $\kappa$-type BEDT-TTF system (BEDT-TTF denotes bis(ethylenedithio)-tetrathiafulvalene), the increase in coupling strength pushes toward the unitary regime of the BCS-BEC crossover.
We elucidate that the identity of the fluctuating superconductivity observed near the Mott boundary is the pseudogap.
Given that the present system is a genuine Fermi liquid showing no other electronic transitions above $T_{\rm c}$, our findings demonstrate that the $\kappa$-type organic system is a suitable research target for a genuinely discussion of the BCS-BEC crossover in strongly correlated electron systems.
We hereafter abbreviate $\kappa$-(BEDT-TTF)$_2$$X$ ($X$ = I$_{3}$, Ag(CN)$_2$H$_2$O, Cu(NCS)$_2$, and Cu[N(CN)$_2$]Br) as $\kappa$-I$_{3}$, $\kappa$-AgCN, $\kappa$-NCS, and $\kappa$-Br, respectively.
The term, d$n$ ($n$=0-8), represents the number of deuterium atoms present in the ethylene group of each BEDT-TTF molecule.
When $n$ is not specified, it indicates that the salt is in its pristine form without any deuteration, namely d0.

Single crystals of the $\kappa$-type salts were synthesized by electrochemical oxidation methods.
Electrical resistance was measured by a standard four-terminal ac method.
Torque magnetometry was carried out using a microcantilever.
Heat capacity measurements were performed using a customized high-resolution calorimeter with single crystals \cite{Imajo2016}.
To determine $H_{\rm c2}$, we performed high-field electrical transport measurements using a 60T pulse magnet.

Figures~\ref{fig2}a and ~\ref{fig2}b present the comparisons of electrical resistivity $R$ and diamagnetic torque d$M_z$/d$H_z$ as a function of temperature (see Appendix for the detailed analysis of d$M_z$/d$H_z$).
Applied fields in the present torque measurements are only a few percent of in-plane $H_{\rm c2}$, and its impact on $T_{\rm c}$ is almost negligible.
We determine $T_{\rm c}$ by the onset temperature of the resistance drop.
High-resolution torque data can provide information on the emergence of fluctuating superconductivity at $T^{\ast}$.
For example, the diamagnetic component of $\kappa$-Br appears below $T^{\ast}$ $\approx$ 18.0~K, which is much higher than $T_{\rm c}$ $\approx$ 11.9~K.
The value of $T^{\ast}$ $\approx$ 18.0~K for $\kappa$-Br is consistent with the temperature at which fluctuating superconductivity emerges \cite{Nam2007,Tsuchiya2013,Dion2009}.
For a full-scale view of the vertical axis, see Appendix.
Similarly, in $\kappa$-d6-Br and $\kappa$-NCS, $T^{\ast}$ observed in this study corresponds to the reported values \cite{Nam2007,Tsuchiya2012,Dion2009,Imajo2022,Matsumura2022} and is significantly higher than $T_{\rm c}$, whereas $\kappa$-AgCN and $\kappa$-I$_{3}$ exhibit $T^{\ast}$ comparable $T_{\rm c}$.
Although fluctuating superconductivity was not observed in $\kappa$-NCS in some previous reports \cite{Nam2007,Furukawa2023}, given the field applied for these measurements and the resolution of the data, these should also be consistent with our present results.
Based on the phase diagram shown in Fig.~\ref{fig1}b, these results imply that strong correlations lead to a wider temperature range of fluctuating superconductivity.

Figures~\ref{fig2}c show the heat capacity data near $T_{\rm c}$ for each salt as $\Delta$$C_{\rm ele}$/$\gamma$$T$ vs. $T$/$T_{\rm c}$ with results reported in Refs. \cite{Matsumura2022,Wosnitza1994,Imajo2018,Imajo2021PRB}.
$\Delta$$C_{\rm ele}$ represents the electronic heat capacity difference between the superconducting and normal states, and $\gamma$ denotes the electronic heat capacity coefficient.
Given the resolution of the present data, the heat capacity is not suitable to detect a small fluctuating contribution and determine $T^{\ast}$.
Correcting the peak broadened by fluctuation effect, the height of the jump at $T_{\rm c}$ in a mean-field approximation $\Delta$$\gamma$/$\gamma$ is evaluated as the dashed curves.
For a $d$-wave superconductor in a weak-coupling limit, 2$\Delta$/$k_{\rm B}$$T_{\rm c}$ $\approx$ 4.3 yields $\Delta$$\gamma$/$\gamma$ $\approx$ 1.0 using a gap function $\Delta$$_0$cos(2$\phi$) in the so-called $\alpha$ model \cite{Padamsee1973}.
The salts with smaller $U$/$W$ exhibit smaller values of $\Delta$$\gamma$/$\gamma$, while those with larger $U$/$W$ show larger values of $\Delta$$\gamma$/$\gamma$.

\begin{figure}
\begin{center}
\includegraphics[width=\hsize]{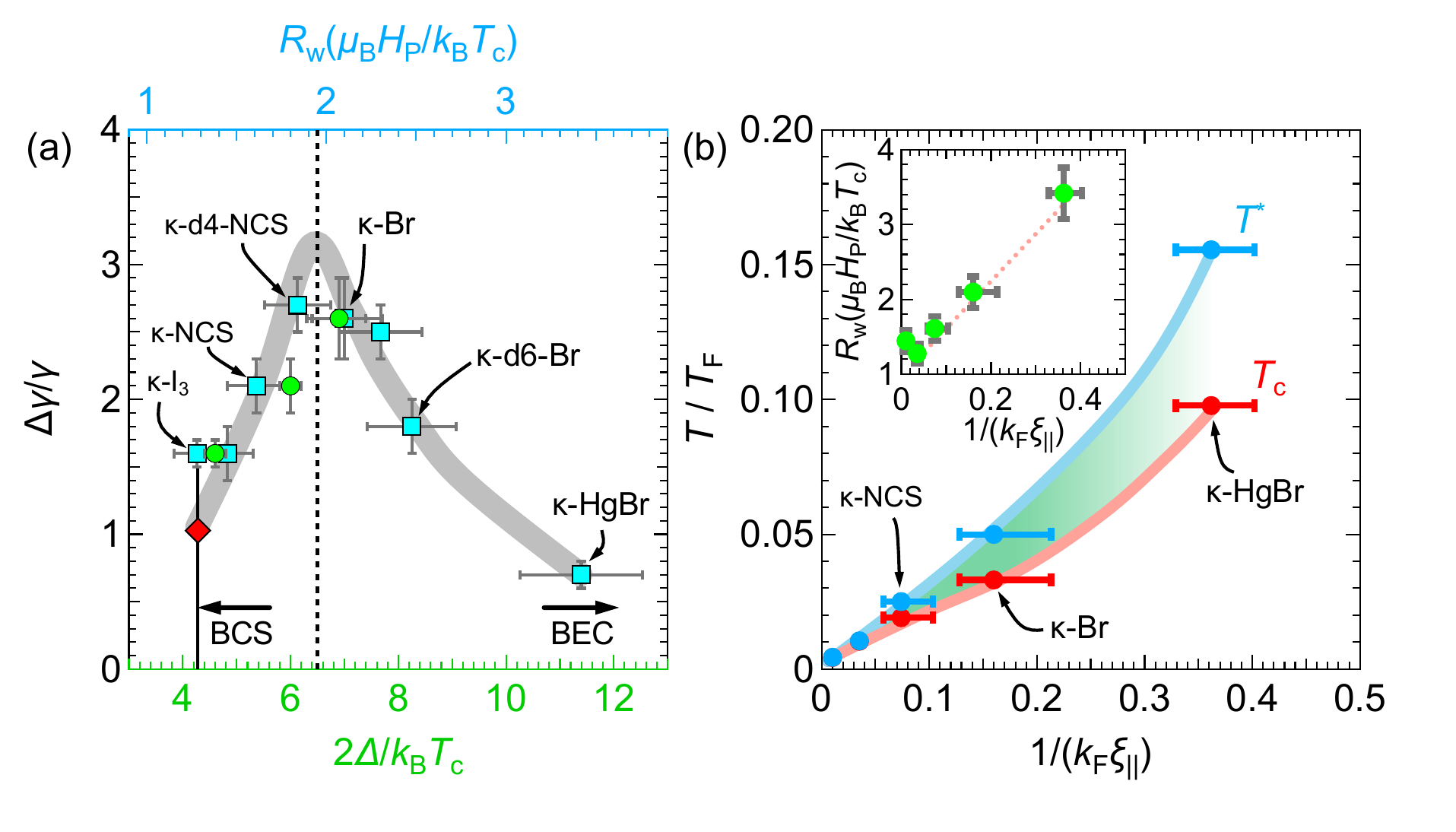}
\end{center}
\caption{
(a) $\Delta$$\gamma$/$\gamma$ versus 2$\Delta$/$k_{\rm B}$$T_{\rm c}$ (bottom axis) and $R_{\rm w}$($\mu$$_{\rm B}$$H_{\rm P}$/$k_{\rm B}$$T_{\rm c}$) (top axis).
The blue and green symbols are plotted to the top and bottom axes, respectively.
The data of $\kappa$-HgBr is taken from Ref.~\cite{Imajo2021PRR}.
The solid and dotted lines indicate the positions of the BCS limit (2$\Delta$/$k_{\rm B}$$T_{\rm c}$ $\approx$ 4.3) and the unitary limit (2$\Delta$/$k_{\rm B}$$T_{\rm c}$ $\approx$ 6.5), respectively.
The red diamond signifies the calculated value in the framework of the $\alpha$ model \cite{Padamsee1973} using a simple weak-coupling $d$-wave gap function $\Delta$$_0$cos(2$\phi$), where $\Delta$$_0$=2.14$k_{\rm B}$$T_{\rm c}$ (2$\Delta$/$k_{\rm B}$$T_{\rm c}$ $\approx$ 4.3).
The gray curve is a visual guide.
(b) 1/$k_{\rm F}$$\xi$$_{\parallel}$ dependence of $T^{\ast}$/$T_{\rm F}$ (blue) and $T_{\rm c}$/$T_{\rm F}$ (red).
The green shaded area corresponds to the pseudogap region.
The inset shows the relation between $R_{\rm w}$($\mu$$_{\rm B}$$H_{\rm P}$/$k_{\rm B}$$T_{\rm c}$) and 1/$k_{\rm F}$$\xi$$_{\parallel}$.
}
\label{fig3}
\end{figure}

The present results indicate that fluctuating superconductivity appears in the temperature range between $T^{\ast}$ and $T_{\rm c}$, which widens in larger $U$/$W$ salts.
We here discuss whether the fluctuating superconductivity corresponds to the formation of the pseudogap.
For opening a pseudogap, the system must be situated in the vicinity of the unitary limit.
At the unitary limit, the pair condensate is optimally reinforced, and therefore, $\Delta$$\gamma$/$\gamma$ becomes maximum \cite{Haussmann2007,Harrison2022,Wyk2016}.
In Fig.~\ref{fig3}a, we show $\Delta$$\gamma$/$\gamma$ for each salt with the horizontal axis of the coupling strength, 2$\Delta$/$k_{\rm B}$$T_{\rm c}$ \cite{Wosnitza1994,Taylor2007,Imajo2018,Imajo2021PRB} for the bottom axis.
We here plot results of prior reports for non-half-filled $\kappa$-type salt $\kappa$-(BEDT-TTF)$_4$Hg$_{2.89}$Br$_8$ ($\kappa$-HgBr)\cite{Imajo2021PRR}, which will be discussed in detail later. 
As the top axis, $R_{\rm w}$($\mu$$_{\rm B}$$H_{\rm P}$/$k_{\rm B}$$T_{\rm c}$) is also shown in this plot, where $R_{\rm w}$, $\mu$$_{\rm B}$, and $H_{\rm P}$ represent the Wilson's ratio, Bohr magneton, and Pauli limit field, respectively.
The values of $R_{\rm w}$ and $H_{\rm P}$ are taken from Refs.~\cite{Imajo2022,Imajo2018,Imajo2021PRB,Wanka1996,McKenzie1999} or determined by the high-field transport shown in Appendix.
Since $H_{\rm P}$ in correlated superconductors is determined by $\Delta$ and $R_{\rm w}$ \cite{Imajo2022,McKenzie1999,Agosta2018}, $R_{\rm w}$($\mu$$_{\rm B}$$H_{\rm P}$/$k_{\rm B}$$T_{\rm c}$) is proportional to the coupling strength.
This plot indicates that $\Delta$$\gamma$/$\gamma$ shows a maximum value around 2$\Delta$/$k_{\rm B}$$T_{\rm c}$ = 6-7.
The value of 2$\Delta$/$k_{\rm B}$$T_{\rm c}$ $\approx$ 6.5 is recently proposed as the magic gap ratio, which can be a thermodynamic indicator of the BCS-BEC crossover that universally observed in various superconductors \cite{Harrison2022}.
This consistency suggests that the unitary region in the $\kappa$-type system also lies close to 2$\Delta$/$k_{\rm B}$$T_{\rm c}$ = 6-7, and that the larger $U$/$W$ salts in proximity to the unitary region exhibit a pseudo gap phase above $T_{\rm c}$.

Although the agreement with the results of the prior studies suggests the BCS-BEC scenario, we must remind that there is much controversy about the presence of the BCS-BEC crossover in cuprates\cite{Harrison2022,Tallon2023,Sous2023}. 
To justify the BCS-BEC scenario in the organic salts, it is crucial to demonstrate that the superconducting phase diagram can be explained by a BCS-BEC crossover phase diagram characterized by 1/($k_{\rm F}$$\xi$).
In Fig.~\ref{fig3}b, we show $T_{\rm c}$ (red) and $T^{\ast}$ (blue) on the $T$/$T_{\rm F}$ vs. 1/($k_{\rm F}$$\xi$$_{\parallel}$) plot.
$T_{\rm F}$ and $\xi$$_{\parallel}$ denote the Fermi temperature and in-plane coherence length, respectively.
The values of $T_{\rm F}$ and $\xi$$_{\parallel}$ are calculated by the superfluid density, heat capacity, and $H_{\rm c2}$ data taken from Refs.~\cite{Wosnitza1994,Imajo2018,Imajo2021PRB,Imajo2021PRR,Wakamatsu2020,Yoneyama2004}.
The inset shows $R_{\rm w}$($\mu$$_{\rm B}$$H_{\rm P}$/$k_{\rm B}$$T_{\rm c}$) vs. 1/($k_{\rm F}$$\xi$$_{\parallel}$).
The positive correlation identifies the relationship between the coupling strength and 1/($k_{\rm F}$$\xi$$_{\parallel}$) \cite{Haussmann2007}.
Figure~\ref{fig3}b indicates that the value of 1/($k_{\rm F}$$\xi$$_{\parallel}$) for $\kappa$-Br is approximately 0.16, which appears smaller than 1 for the unitary regime but not negligible.
In the BCS-BEC framework, $T_{\rm c}$/$T_{\rm F}$ is approximately 0.2 in the BEC side.
However, in the case of $\kappa$-Br, $T_{\rm c}$/$T_{\rm F}$$\sim$0.03, which is much smaller than 0.2.
It is worth noting that the BCS-BEC phase diagram of Fermi gases slightly differs from that of superconductors due to the kinetic energy degrees of freedom associated with the electron motion in a periodic lattice.
For two-dimensional superconductors, $T_{\rm c}$/$T_{\rm F}$ in the BEC limit decreases and approaches approximately 1/8 \cite{Harza2019}.
Moreover, $T_{\rm c}$/$T_{\rm F}$ depends on the effective mass of fermions and pairing symmetry.
In cases where the effective mass becomes heavier or the symmetry is $d$-wave, changes in the BCS-BEC phase diagram, such as a decrease in $T_{\rm c}$/$T_{\rm F}$, occur \cite{Melo1993,Randeria2014,Chen2022}.
Given the $d$-wave superconducting state and the strongly renormalized effective mass due to large $U$/$W$ in $\kappa$-Br, $T_{\rm c}$/$T_{\rm F}$$\sim$0.03 is reasonable.
In the case of Li$_x$ZrNCl gate-controlled layered superconductors \cite{Nakagawa2021}, pseudogap formation occurs when the Li content $x$ is below 0.05.
At $x$ $\approx$ 0.05, the values of 1/($k_{\rm F}$$\xi$) $\approx$ 0.1 and $T_{\rm c}$/$T_{\rm F}$ $\approx$ 0.05 yield $T^{\ast}$/$T_{\rm c}$ $\approx$ 1.5.
These values are consistent with the results obtained for $\kappa$-Br, which suggests that fluctuating superconductivity in the $\kappa$-type system can also be identified as a pseudogap phase.

We note that superconducting properties of $\kappa$-HgBr (1/($k_{\rm F}$$\xi$$_{\parallel}$)=0.3-0.4 and $T_{\rm c}$/$T_{\rm F}$$\approx$0.1) \cite{Oike2017,Suzuki2022} can also be explained in the framework expected from extrapolations of $\kappa$-Br and $\kappa$-NCS.
$\kappa$-HgBr has been considered as a unique salt that does not share the same electronic phase diagram as Fig.~\ref{fig1}b due to its distinct features, such as band filling deviating from half and quantum-spin-liquid-like behavior\cite{Oike2017}.
The previous studies \cite{Oike2017,Suzuki2022} have discussed the BCS-BEC crossover of $\kappa$-HgBr as distinct from superconducting properties of half-filled $\kappa$-type salts.
However, the present results demonstrate that $\kappa$-HgBr also shares the same superconducting phase diagram as the other $\kappa$-type superconducting state from the perspective of the BCS-BEC physics.

\begin{figure}
\begin{center}
\includegraphics[width=\hsize]{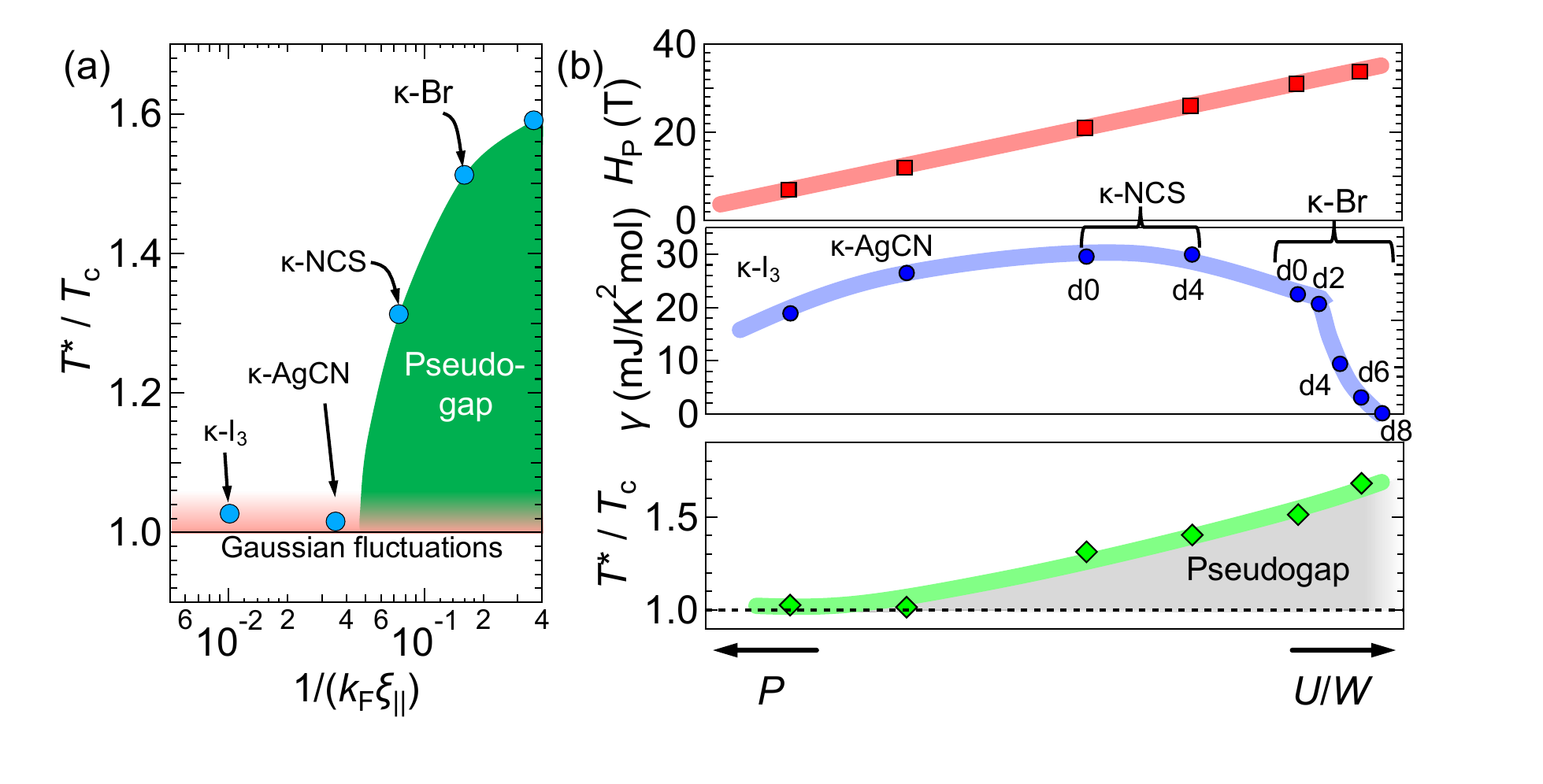}
\end{center}
\caption{
(a) Evolution of the ratio $T^{\ast}$/$T_{\rm c}$ depending on 1/$k_{\rm F}$$\xi$$_{\parallel}$.
The Gaussian-fluctuation region (red shaded area) is limited to only $T^{\ast}$/$T_{\rm c}$ $\approx$ 1.05-1.1 or less, while the pseudogap phase (green area) covers the higher-temperature region even above $T^{\ast}$/$T_{\rm c}$ $>$ 1.1 when 1/($k_{\rm F}$$\xi$$_{\parallel}$) $>$ 0.05.
(b) Shcematic $U$/$W$ dependence of $H_{\rm P}$ (top), $\gamma$ (middle), and $T^{\ast}$/$T_{\rm c}$ (bottom) of $\kappa$-type salts.
The curves are visual guides.
}
\label{fig4}
\end{figure}

These consideration unveil the BCS-BEC physics in the $\kappa$-type superconductors.
Nevertheless, we need to evaluate the effect of thermal fluctuations on the present superconductivity.
To make the fluctuating region clearer, in Fig.~\ref{fig4}a, we show $T^{\ast}$/$T_{\rm c}$ vs. 1/($k_{\rm F}$$\xi$$_{\parallel}$) as a semilogarithmic plot.
In the classical framework, order parameters are fluctuated by thermal energy, and the behavior in the critical region is described by Gaussian fluctuations.
The Gaussian critical region $T_{\rm G}$/$T_{\rm c}$ is estimated by the equation $T_{\rm G}$/$T_{\rm c}$ =1+2($k_{\rm B}$/8$\pi$$\Delta$$C$$\xi$$^3$)$^2$, where $\Delta$$C$ denotes heat capacity jump at $T_{\rm c}$ \cite{Quader1988}.
$T_{\rm G}$/$T_{\rm c}$ for the present $\kappa$-type salts is in the range of 1.05-1.1, which corresponds to the broadened region of the heat capacity data shown in Fig.~\ref{fig2}c.
Nevertheless, it is hard to reconcile the Gaussian-fluctuation region with the value of $T^{\ast}$/$T_{\rm c}$ when 1/($k_{\rm F}$$\xi$$_{\parallel}$) $>$ 0.05, as shown in Fig.~\ref{fig4}a.
Therefore, the wide fluctuating region for larger $U$/$W$ salts originates from the formation of a pseudogap.

Finally, we consider electron-correlation dependence of $\kappa$-type superconductivity.
The exact calculation of the values of low-temperature $U$/$W$ is challenging due to the lack of the structural data at low temperatures near $T_{\rm c}$.
To estimate the relative magnitude of $U$/$W$ roughly,  we here use $H_{\rm P}$, which is strongly influenced by $U$/$W$.
In the top panel of Fig.~\ref{fig4}b, we show the relative positions of the measured salts, to make the $U$/$W$ dependence of $H_{\rm P}$ linear in this plot.
Long-standing experimental studies \cite{Kanoda2006,Nakazawa2018} have established the $U$/$W$ dependence of $\gamma$, as shown in the middle panel of Fig.~\ref{fig4}b.
In the low $U$/$W$ region, $\gamma$ increases with increasing $U$/$W$.
This behavior can be understood within the Brinkman-Rice framework \cite{Brinkman1970}, which suggests that electron correlations result in the renormalization of the electron mass.
However, in the immediate vicinity of the Mott boundary, a significant reduction in $\gamma$ is observed for $\kappa$-d$n$-Br ($n$ = 4-8) \cite{Nakazawa2018,Matsumura2022}.
This is because the Mott transition is first-order, and the normal state is eroded with the insulating state by inhomogeneity, leading to the reduction in $\gamma$.
Therefore, the abrupt drop in $\gamma$ should not relate to the BCS-BEC crossover.
On the other hand, the gradual change in $\gamma$, observed in the region where the pseudogap opens (Fig.~\ref{fig4}b and c), might be caused by pairing fluctuations in the normal state in the BEC region \cite{Harrison2022,Wyk2016}.
To understand the influence of the BCS-BEC crossover on the normal state, further future studies are required.

Our comprehensive investigation of the superconducting state in the $\kappa$-type organic system revealed the formation of a pseudogap near the unitary limit through the tuning of electron correlations.
Given that the normal state of the $\kappa$-type salts is a simple Fermi liquid without any other electronic ordering, the present findings are crucial for identifying the essential parameters required to gain an accurate understanding of BCS-BEC physics.
It is worth noting that the crossover behavior of this system can be controlled through external pressure.
Hence, it is desirable to conduct detailed high-pressure studies of the $\kappa$-type system that focus on the pseudogap in the future.

This study was partly supported by JSPS KAKENHI Grant (20K14406, 22H04466).

\renewcommand{\thefigure}{S\arabic{figure}}
\clearpage
\onecolumngrid
\appendix
\begin{center}
\large{\bf{Supplementary Materials for\\
Pseudogap formation in organic superconductors
}}
\end{center}

\section{Evaluation of diamagnetic signal derived from magnetic torque measurements}
\begin{figure}[hh]
\begin{center}
\includegraphics[width=0.5\linewidth,clip]{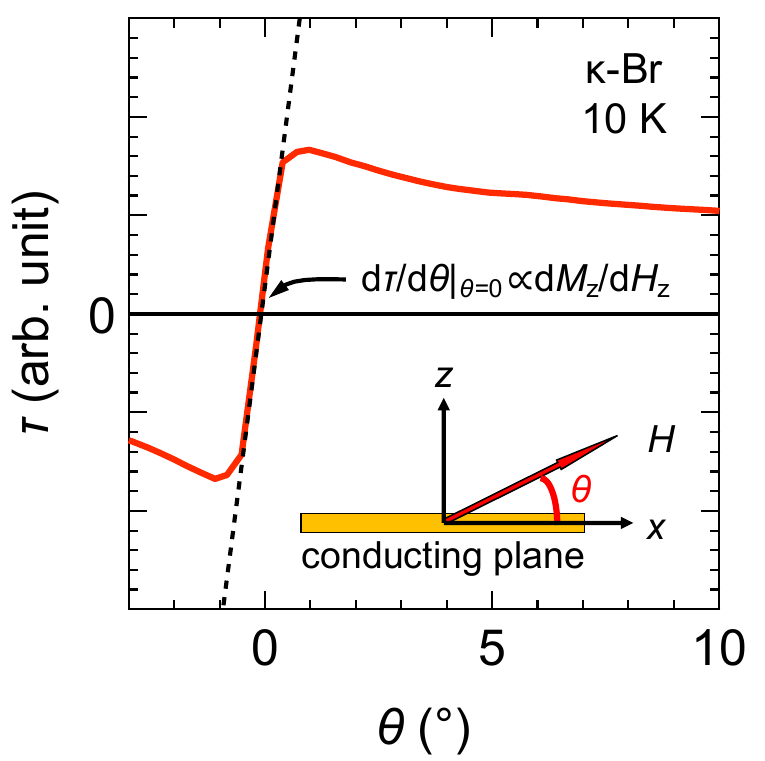}
\end{center}
\caption{
A typical example of angle dependence of magnetic torque for $\kappa$-type superconductors.
The inset illustrates the definition of angle and direction in the present measurements.
}
\label{fig5}
\end{figure}

In Fig.~2 of the main text, we present the temperature dependence of the perpendicular component of the diamagnetic susceptibility of superconductivity, d$M_z$/d$H_z$.
To obtaine d$M_z$/d$H_z$, we employed angle-resolved magnetic torque measurements using microcantilevers.
As shown in Fig.~\ref{fig5}, we measured the angle-dependent torque of each sample in a magnetic field at various temperatures.
The inset displays a schematic of the sample setup for the measurements, with the direction parallel to the in-plane direction set as 0 degrees.
In the angle range of $\pm$0.5 degrees, the torque signal can be described by the following formula,
\begin{equation}
\frac{d\tau}{d\theta}|_{\theta=0} = -\mu_0H_{x} \frac{d(M\times H)}{dH_{z}} |_{\theta=0} = \mu_0H_{x} \frac{d(H_{x}M_{z}-M_{x}H_{z})}{dH_{z}} |_{\theta=0} = \mu_0H_{x} \frac{d[H_{x}H_{z}(M_{z}/H_{z}-M_{x}/H_{x})]}{dH_{z}} |_{\theta=0}.
\end{equation}
Here, we use the relations, $H_x$=$H$cos$\theta$ and $H_z$=$H$sin$\theta$.
In the mixed state of two-dimensional superconductors, since $|$$M_z$/$H_z$$|$ is much larger than $|$$M_x$/$H_x$$|$ in this angle range, the formula can be rewritten as 
\begin{equation}
\frac{d\tau}{d\theta}|_{\theta=0} \approx -\mu_0H_{x}^2 \frac{dM_{z}}{dH_{z}} |_{\theta=0} \propto \frac{dM_{z}}{dH_{z}} |_{\theta=0}
\end{equation}
in a constant field.
Therefore, the slope of the angle-dependent torque near the in-plane configuration is proportional to the perpendicular component of the diamagnetic susceptibility of superconductivity.

\newpage
\section{Temperature dependence of $d$$M_z$/$d$$H_z$}
\begin{figure*}[hh]
\begin{center}
\includegraphics[width=\linewidth,clip]{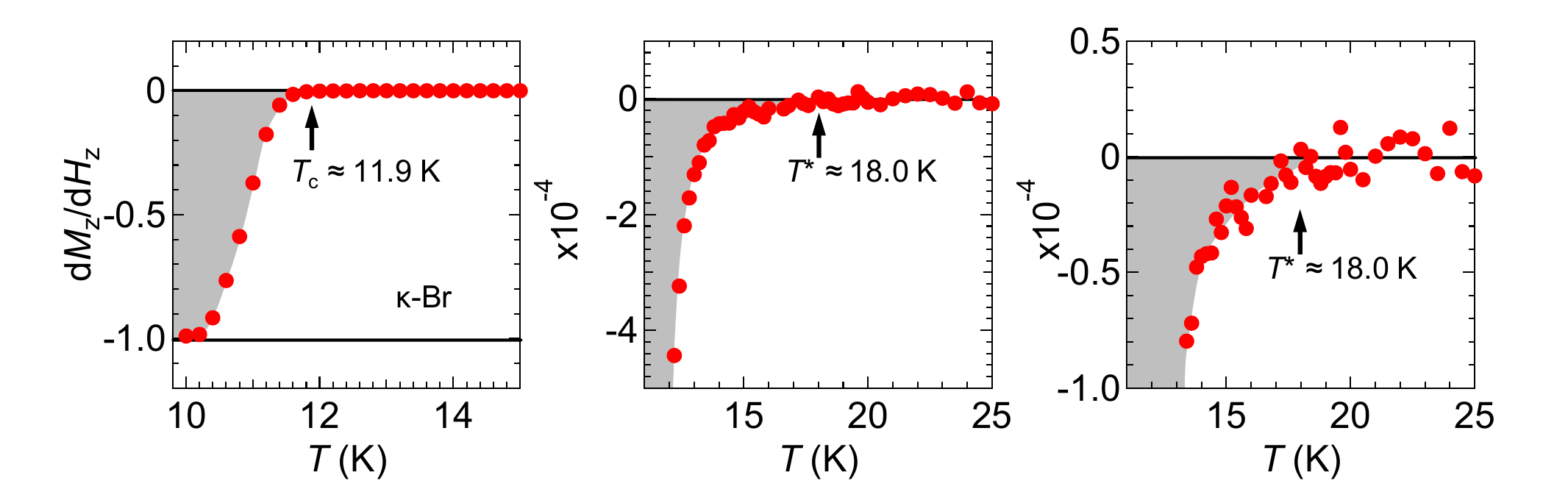}
\end{center}
\caption{Temperature dependence of d$M_z$/d$H_z$ of $\kappa$-Br.
Shaded areas indicate the diamagnetic component.
}
\label{fig6}
\end{figure*}

In Fig. 2b in the main text, we show enlarged plots of d$M_z$/d$H_z$ to emphasize the emergence of the diamagnetic component below $T^{\ast}$.
In Fig.~\ref{fig6}, we plot d$M_z$/d$H_z$ vs. $T$ of $\kappa$-Br at various scales.
From the left full-scale view, the onset temperature of the emergence of the diamagnetic component appears to be more $T_{\rm c}$ than $T^{\ast}$, which corresponds to the resistivity data shown in Fig.~2a.
Nevertheless, from the center and right figures, which are enlargements of the left figure, we can find that the diamagnetic component is present even above $T_{\rm c}$, as discussed in the main text.
Thus, the fluctuation (pseudogap) region between $T_{\rm c}$ and $T^{\ast}$ can be determined from the difference between the resistivity and torque data, respectively.

\newpage
\section{High-field measurement for determining $H_{\rm c2}$}
\begin{figure}[hh]
\begin{center}
\includegraphics[width=0.7\linewidth,clip]{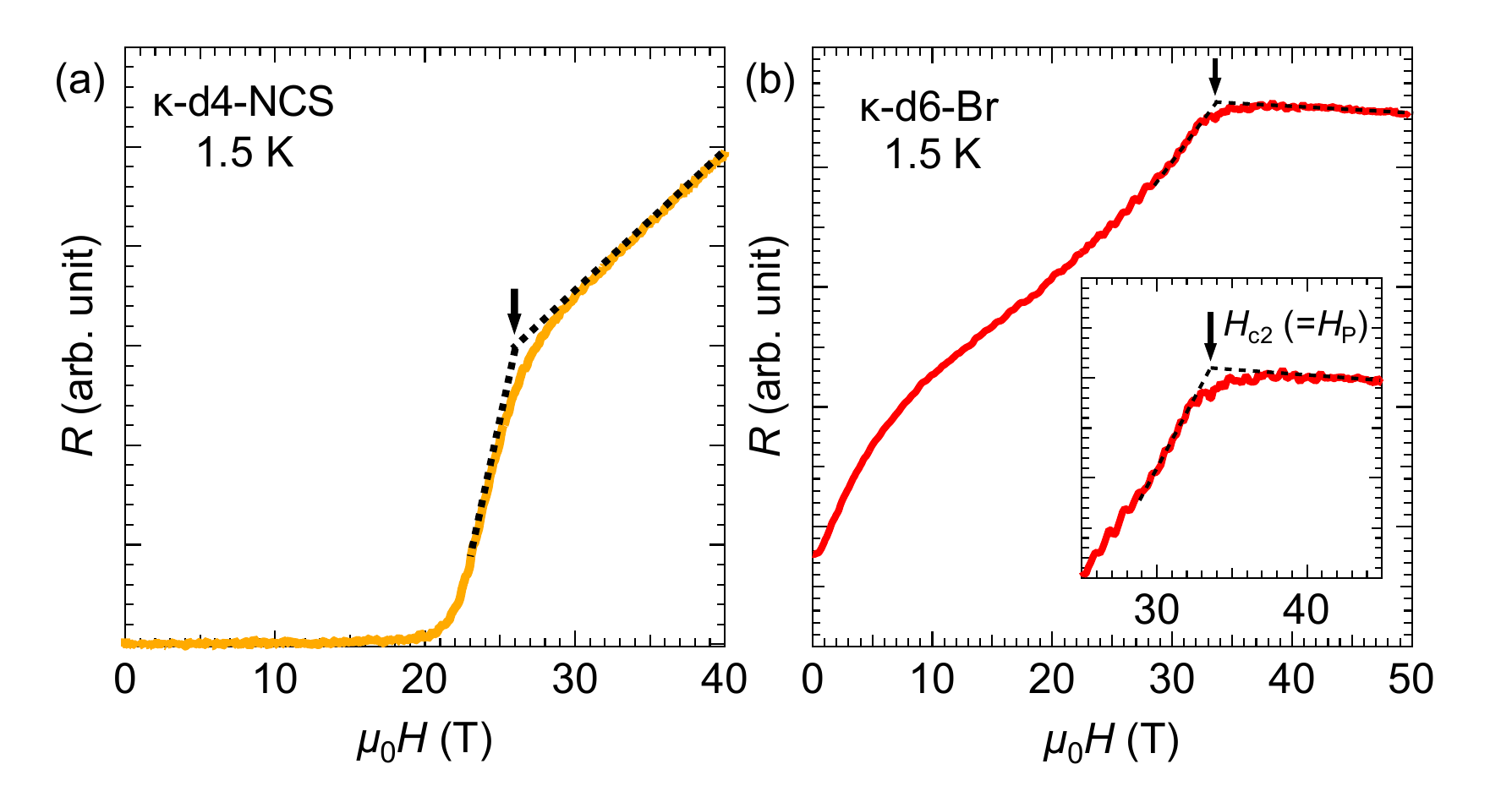}
\end{center}
\caption{
Magnetoresistance of (a) $\kappa$-d4-NCS and (b)  $\kappa$-d6-Br in an in-plane field at 1.5~K.
Arrow indicates $H_{\rm c2}$, which corresponds to $H_{\rm P}$.
The inset of (b) shows an enlarged view around $H_{\rm c2}$.
}
\label{fig7}
\end{figure}

 In the main text, we evaluate the coupling strength using the dimensionless ratio $R_{\rm w}$($\mu$$_{\rm B}$$H_{\rm P}$/$k_{\rm B}$$T_{\rm c}$), as shown in Fig. 3a, as $H_{\rm P}$ in correlated superconductors is determined by $\Delta$ and $R_{\rm w}$~\cite{McKenzie1999,Agosta2018}.
 For $\kappa$-Br, $\kappa$-NCS, $\kappa$-AgCN, $\kappa$-I$_3$, and $\kappa$-HgBr, $H_{\rm P}$ was reported in Refs.~\cite{Imajo2018,Imajo2022,Imajo2021PRB,Wanka1996,Imajo2021PRR}.
 In this study, we obtained $H_{\rm P}$ of $\kappa$-d4-NCS and $\kappa$-d6-Br using a high-field transport measurement with a 60T pulse magnet.
 Figure~\ref{fig7} displays the in-plane magnetic field dependence of the electrical resistance of $\kappa$-d4-NCS and $\kappa$-d6-Br at 1.5~K.
 In the case of these salts, $H_{\rm P}$ corresponds to $H_{\rm c2}$, as the orbital pair breaking effect is quenched when a field is applied parallel to the two-dimensional conducting plane.
 Since we determine $T_{\rm c}$ from the onset of the resistance drop, as shown in Fig.~2a, $H_{\rm c2}$ is also determined by the onset position indicated by the arrow.

\end{document}